\documentclass[prl,twocolumn,showpacs]{revtex4}
\usepackage{amsmath}

\newcommand{\ket}[1]{\left| #1\right>}
\begin{document}
\title{Self-Interacting Dark Matter with Flavor Mixing }
\author{Mikhail V. \surname{Medvedev}}
\email{medvedev@cita.utoronto.ca}
\homepage{http://www.cita.utoronto.ca/~medvedev}
\altaffiliation[Also at the ]{Institute for Nuclear Fusion,
Russian Research Center ``Kurchatov Institute", Moscow 123182, Russia}
\affiliation{Canadian Institute for Theoretical Astrophysics, 
University of Toronto, Toronto, Ontario, Canada, M5S 3H8}

\begin{abstract}
The crisis of both cold and collisional dark matter (DM) models is
resolved by postulating flavor mixing of DM particles. Flavor-mixed 
particles segregate in the gravitational field to form dark halos 
composed of heavy mass eigenstates. Since these particles are mixed 
in the interaction basis, elastic collisions convert some of heavy 
eigenstates into light ones which leave the halo. 
This annihilation-like process will soften dense central cusps of halos. 
The proposed model accumulates most of the attractive features of 
self-interacting and annihilating DM models, but does not suffer from 
their severe drawbacks. This model is natural does not require fine tuning.
\end{abstract}
\maketitle

Dark matter constitutes most of the mass in the Universe, but its nature
and properties remain largely unknown, except, perhaps, it is cold
(at the time of decoupling, i.e., massive) and dark (weakly interacting).
Recent observations indicate a failure of the conventional cold dark matter
(CDM) paradigm on small (galactic) scales. The resolution of the discrepancy 
between theory and observations calls for new ideas and should allow to further 
constrain properties and nature of otherwise undetectable dark matter.

A model of structure formation in a universe with CDM is in excellent agreement 
with observations on large scales ($\gg$~Mpc) \cite{Bahcall+99} which, thus,
supports the hypothesis that dark matter particles are heavy ($\gtrsim$~GeV) and
weakly interacting with baryonic matter and photons. Recent observations on 
galactic and sub-galactic scales appear to be in conflict with the CDM model,
which predicts singular ($\rho\sim1/r$ or even steeper) central density cusps
\cite{Navarro+97,Klypin+99,Moore+99a} and a large number of low-mass halos
\cite{Klypin+99,Moore+99b}. In contrast, observations indicate low-density
cores with shallow profiles \cite{Moore94,DM97,DH00} and a smaller number 
of satellites \cite{M98}. Also, some observations seem to indicate the 
invariance of the core density with respect to the mass of the halo 
\cite{Firmani+00}. 

A number of models have recently been proposed in order to explain the 
discrepancy. Some of them, such as the models of self-interacting, 
warm, annihilating, and decaying dark matter 
\cite{SS00,Hannestad+00,Kaplinghat+00,Cen00} are just (slight) modifications 
of the standard CDM paradigm, while others, e.g., the fluid, repulsive, fuzzy, 
bosonic, mirror, etc. dark matter models 
\cite{Peebles00,Goodman00,Hu+00,RiottoTkachev00,MT00} are rather exotic. 
Nearly all theories may be tuned to solve the problem, they may be 
motivated from particle physics, and, finally, they cannot be ruled 
out by present observational data.

The fact that many predictions of the CDM cosmological models are in excellent
agreement with a wide variety of observational results, from the microwave 
background fluctuations to the large-scale structure, suggests that the 
CDM model is a good first approximation, but it has to be corrected on a
small (galactic) scale. Among many possibilities, the simplest one is to 
assume that weakly interacting CDM particles are, in fact, strongly
self-interacting \cite{SS00,Carlson+94,deLaix+95}. Evolution of the 
structure in the presence of self-interacting dark matter (SIDM) indeed 
results in a smaller number of satellites and formation of constant 
density cores within a dynamical time-scale modified by collisions 
\cite{Burkert00,Moore+00,Yoshida+00,KochanekWhite00,Dave+00}. 
On a longer time-scale, however, a SIDM halo is unstable to core collapse, 
which leads to the formation of a steep $\sim1/r^2$ density cusp, making 
the problem even worse \cite{KochanekWhite00} (but see Ref. \cite{Dave+00}).
The detection of elongated cores in clusters via lensing observations 
\cite{ME00} is inconsistent with the simple SIDM model.
Thus, it seems that such an one-component model is an over-simplification,
because, in general, dark matter particles may be flavor mixed, that is their 
interaction (or flavor) eigenstates are superpositions of mass eigenstates. 

In this paper we study the properties of dark matter halos in the model of
the SIDM with flavor mixing. We demonstrate that {\em elastic} scattering
of such dark matter particles results in a conversion of mass eigenstates and 
calculate the cross-section of this process. The conversion process will lower 
the density in the core region and prevent the cusp formation. This is somewhat
similar to the annihilating dark matter model \cite{Kaplinghat+00}, but
our model does not suffer from the ``annihilation catastrophe'' in the 
early Universe because of the dynamical equilibrium of the opposite 
conversion processes in a homogeneous medium. 

For the sake of simplicity, let us assume that there exist two flavors,
the strongly self-interacting, $\ket{\cal I}$, and the non-self-interacting
or ``sterile'', $\ket{\cal S}$, ones. These interaction eigenstates are
a superposition of two mass eigenstates, the ``heavy'', $\ket{m_h}$, 
and the ``light'', $\ket{m_l}$, ones. We write 
\begin{equation}
{\ket{m_h}\choose{\ket{m_l}}}=
\left(\begin{array}{cc}
\cos\vartheta & \sin\vartheta\\
-\sin\vartheta & \cos\vartheta
\end{array}\right)
{\ket{\cal I}\choose{\ket{\cal S}}},
\label{mix}
\end{equation}
where $\vartheta$ is the mixing angle. Throughout the paper, we assume for 
simplicity that $m_h\gg m_l$, so that heavy states are {\em non-relativistic} 
and light states are {\em relativistic}. Generalizations of the above 
simplifications are straightforward. We should note that, in general, the 
treatment of non-relativistic species requires the wave packet approach 
\cite{Dolgov}, and not the plane wave one used above. In some instances, this 
may result in measurable corrections \cite{GKL91,GKL92}. Here we neglect by 
these complications and use the standard plane wave approach, keeping in mind
that a particle is a localized in space wave packet with the width $\delta x$
determined by the production process.

The concept of flavor eigenstates arises when interactions of particles
are considered. In the field-free theory the particle fields in the mass
basis have a physical meaning instead. In general, such mass eigenstates have 
different momenta and energies, $E^2_{h,l}=|{\bf p}_{h,l}|^2c^2+m^2_{h,l}c^4$. 
Between the interactions, the mass eigenstates propagate independently, with 
different velocities ${\bf v}_{h,l}={\bf p}_{h,l}c^2/E_{h,l}$. Thus at times
\begin{equation}
t\gg t_s\sim\delta x/|{\bf v}_l-{\bf v}_h|\sim\delta x/c
\end{equation}
these states are separated from each other and their wave functions 
no longer overlap. The separation time above is negligibly small
compared to a galactic dynamical time scale; thus, this process is essentially 
instantaneous. In a gravitational field different eigenstates segregate by mass.
Non-relativistic $\ket{m_h}$-states form halos and the large-scale structure,
they represent the conventional cold dark matter. Relativistic 
$\ket{m_l}$-states correspond to the ``hot'' dark matter component which 
cannot form structures. This component has a much smaller total mass, however, 
(the number density of both mass states are equal by symmetry) and, hence, 
is dynamically unimportant.

As dark matter halos form, the density in the central parts increases
and, at some point, self-interactions of the dark matter particles become
important. Let us consider the elementary act of {\em elastic} scattering
of two particles, i.e., $\ket{m_h}$ eigenstates. The initial wave
function of two interacting particles in the center of mass frame, 
according to equation (\ref{mix}), is
\begin{equation}
\Psi_i=\left(e^{ikz}\pm e^{-ikz}\right)\ket{m_h}
=e^{\pm ikz}\left(\cos\vartheta\ket{\cal I}+\sin\vartheta\ket{\cal S}\right),
\end{equation}
where ``$+$'' corresponds to $\Psi_i$ symmetric to interchange of particles 
(integer total spin) and ``$-$'' -- to an antisymmetric $\Psi_i$ (half-integer 
total spin), the exponents represent two waves, propagating to the right, 
and to the left, and $e^{\pm ikz}\equiv\left(e^{ikz}\pm e^{-ikz}\right)$ 
is the short-hand notation. For scattering, the interaction basis is 
appropriate, rather than the mass basis, hence the expansion above. During 
the scattering event, only $\ket{\cal I}$-component is changing, since 
$\ket{\cal S}$-component does not interact. 
The wave function at large distances after scattering thus becomes
\begin{eqnarray}
\Psi_s&\approx&\left(e^{\pm ikz}+\frac{f_\pm(\theta)}{r}\,e^{ikr}\right)
\cos\vartheta\ket{\cal I}+e^{\pm ikz}\sin\vartheta\ket{\cal S}
\nonumber\\
&=&\left(e^{\pm ikz}+\cos^2\vartheta\,\frac{f_\pm(\theta)}{r}\,e^{ikr}\right)
\ket{m_h}  \nonumber\\
& &{ }-\cos\vartheta\sin\vartheta\,\frac{f_\pm(\theta)}{r}\,e^{ikr}\ket{m_l},
\end{eqnarray}
where the combination $f_\pm(\theta)=f(\theta)\pm f(\pi-\theta)$ arises 
because particles are indistinguishable and  $f(\theta)$ is the amplitude of 
scattering of flavor states. The radial part of the wave function represents a 
diverging scattered wave. One can clearly see that an initial heavy eigenstate 
acquires, upon scattering, a light eigenstate admixture. In other words,
a heavy eigenstate may be converted into a light eigenstate.
The differential cross-section of the conversion process is the 
scattering amplitude squared, that is 
\begin{subequations}
\begin{equation}
d\sigma_{c,\pm}=\left|\cos\vartheta\sin\vartheta\, 
f_\pm(\theta)\right|^2\, d\Omega,
\end{equation}
where $d\Omega=2\pi\sin\theta\, d\theta$ is the solid angle.
The cross-section of pure scattering is 
\begin{equation}
d\sigma_{s,\pm}=\left|\cos^2\vartheta\, f_\pm(\theta)\right|^2\, d\Omega.
\end{equation}
\end{subequations}
Of course, energy and momentum must be conserved in these processes. 
Introducing the integral cross-section as 
$\sigma_\pm=\int\left|f_\pm(\theta)\right|^2d\Omega$, we write
\begin{equation}
\sigma_{c,\pm}=\sin^2\vartheta\cos^2\vartheta\, \sigma_\pm, \qquad
\sigma_{s,\pm}=\cos^4\vartheta\, \sigma_\pm.
\end{equation}
It is assumed here that the interacting particles of spin $s$  have a 
particular value of their total spin. If it is not so, one has to average 
over all spin states, which yields \cite{LL3} for the conversion cross-section:
\begin{equation}
\sigma_{c}=\frac{s}{2s+1}\sigma_{c,+}+\frac{s+1}{2s+1}\sigma_{c,-}
\end{equation}
and similarly for the scattering cross-section $\sigma_s$. The cross-sections 
for the light mass eigenstate interactions may be obtained similarly.

Let us now consider a dark matter halo. When dark matter particles 
(heavy mass eigenstates) interact with each other, they can either 
scatter with some cross-section $\sigma_s$ or be converted into light mass 
eigenstates with the cross-section 
\begin{equation}
\sigma_c=\tan^2\vartheta\,\sigma_s.
\end{equation}
Since these light eigenstates are relativistic, 
the conversion process is analogous to the annihilation of dark matter 
particles. Since the emergent light eigenstates are relativistic,
they leave the dark halo, thus, decreasing its density and mass.
If the mixing angle is not too small, $\vartheta\sim1$, collisional heat 
transport, which flattens the core region by making it nearly isothermal, 
and dark matter ``annihilation'' (conversion) become important at roughly 
the same density of particles and, hence, operate simultaneously. 
Hence, on the onset of the core collapse, when the infall velocity is small,
there is enough time to annihilate the density enhancement and prevent 
formation of a central cusp. Only direct particle simulations
are capable of determining what the spatial structure of the dark matter 
halo will be in this regime.

As one can see, most of the assumptions adopted in the beginning are
not crucial for our model. The model requires only that the interaction
strengths of flavor eigenstates and the masses of mass eigenstates
be (significantly) different. No fine tuning is necessary.

A remarkable difference of the eigenstate conversion process from pure
annihilation is that particles are not permanently destroyed via conversion.
This is a reversible process: $\ket{m_h}$ eigenstates are ``regenerated''
in interactions of $\ket{m_l}$ eigenstates. By symmetry, the conversion 
cross-sections of regeneration and annihilation are identical. Now consider 
the early Universe. The density fluctuations in it are negligible. Therefore,
both processes occur at the same rate. They balance each other exactly,
no matter how large the particle density is. Thus, no excess of particles
of one type over particles of another type is produced. In contrast, 
conventional annihilation is extremely rapid at high densities. The 
annihilating dark matter model \cite{Kaplinghat+00}, thus, suffers from
the ``annihilation catastrophe''. Namely, dark matter must be destroyed 
in the early Universe epoch, long before the large scale structure began 
to form.

We also should note that the conversion and pure annihilation cross-sections
have, in general, different dependence on the relative velocity, $v$, of
interacting particles. In the absence of long-range interactions and
in the non-relativistic limit, the annihilation cross-section scales
usually as $\sigma_a\propto v^{-1}$, while $\sigma_c$, being proportional
to the cross-section of elastic scattering, does not depend on $v$, i.e., 
$\sigma_c\propto v^0$. These scalings are typical but not universal, 
however, and depend on the interaction model at hand 
\cite{Kaplinghat+00,Hogan+00,Wyithe+00}.

The dynamics of SIDM model with flavor mixing is quite rich. In the limit of
no mixing, $\vartheta=0$, the model reduces to the standard SIDM model. 
If the mixing angle is rather small, elastic collisions rapidly establish 
isothermal equilibrium throughout a dark matter halo. The halo becomes
relatively spherical in this case. Moreover, its core will collapse.
Particle conversions will be important deep inside the collapsed core
where the density is high enough. Finally, if $\vartheta\gtrsim1$, 
conversions of dark matter particle rapidly lower the halo density.
An almost spherical isothermal core never forms. The detection of 
elongated shapes from lensing observations of clusters \cite{ME00} 
seems to favor the last scenario.

To summarize, we presented a model of self-interacting dark matter with 
flavor mixing. This model is well motivated from particle physics
(e.g., mixing of quarks, neutrinos, etc.) within the standard model, 
it is robust and does not require fine tuning of any kind. We have 
demonstrated that dark matter halos are composed of particle mass
eigenstates. Being mixed in the interaction basis, some of heavy 
eigenstates are converted in elastic collisions into light eigenstates
and leave the halo, in a striking analogy with halo annihilation.
Accumulating many attractive features of SIDM and annihilating dark
matter models (e.g., formation of soft cores and a fewer number of sub-halos,
while preserving CDM properties on a large scale), our model does not suffer 
from their caveats, such as the core collapse induced $r^{-2}$-cusps
and too spherical cores in the SIDM model, and the catastrophically 
strong annihilation of dark matter in the dense early Universe.

The author is grateful to Jim Peebles for interesting and insightful 
discussions and his interest in this work and to CITA for the
financial support.



\end{document}